# The Physics of Quantum 2.0: Challenges in understanding Quantum Matter

*Siddhartha Lal [1*]*
[1] Department of Physical Sciences, Indian Institute of Science Education and Research Kolkata, Mohanpur Campus, West Bengal 741246, India
Email: slal@iiserkol.ac.in
* Author to whom any correspondence should be addressed.

*Mayank Shreshtha[2]*
[2] Cavendish Laboratory, Department of Physics, University of Cambridge, Cambridge CB3 0HE, U.K.
Email: ms3103@cam.ac.uk

**Abstract**
Almost a century on from the culmination of the first revolution in quantum physics, we are poised for another. Even as we engage in the creation of impactful quantum technologies, it is imperative for us to face the challenges in understanding the phenomenology of various emergent forms of quantum matter. This will involve building on decades of progress in quantum condensed matter physics, and going beyond the well-established Ginzburg-Landau-Wilson paradigm for quantum matter. We outline and discuss several outstanding challenges, including the need to explore and identify the organisational principles that can guide the development of theories, key experimental phenomenologies that continue to confound, and the formulation of methods that enable progress. These efforts will enable the prediction of new quantum materials whose properties facilitate the creation of next generation technologies.

**The journey thus far.**

Leaps in technology arise typically from underlying progress in fundamental science. The truth of this statement can be learnt in the evolution of condensed matter physics as a discipline. As an example, consider the case of superconductivity. The seemingly serendipitous discovery of a state of matter with zero resistance in mercury by Kamerlingh Onnes and his team at Leiden in 1911 certainly qualifies as a eureka moment. The credit for this lies, however, with another momentous achievement by the same pioneering team of low-temperature physicists three years earlier in the guise of the first successful liquefaction of helium. Indeed, it appears that the goal of the 1911 experiment was to test a transfer system for liquid helium into a cryostat, with a delightful spin-off in the availability of electrical measurements of metals at temperatures as low as 1 Kelvin. This highlights the breakthrough discovery of superconductivity as an outcome of a carefully constructed research programme in cryogenics [1]. Much has happened in the field of condensed matter physics since then, as can be glimpsed in a rough map of the field over most of the 19[th] and 20[th] centuries that is offered in Figure 1.

Today, we are poised on the threshold of the creation of impactful quantum technologies for communication, sensing and computation. Naturally, our expectations for these technologies are rooted in the understanding of how matter is governed by the guiding principles of quantum physics, and how key quantum properties can be manifested at the macroscale to facilitate devices based on the desired functionality of materials. This ambitious goal lies at the heart of quantum condensed matter physics and its interface with materials science. Building on the quantum revolution of the first half of the 20[th] century, early successes in this venture involved understanding why metals are different from insulators, how magnets and superconductors arise respectively from electronic correlations and electron-phonon interactions etc. These helped in developing a language for the emergence of novel phases of matter that cannot be attributed merely to the properties of their constituents. They also led to the creation of vital technologies, e.g., superconducting magnets essential to magnetic resonance imaging machines as well as particle accelerators. Equally important was understanding the motion of electrons in semiconductors, as this enabled the translation of the first transistors into the circuitry at the heart of all modern technologies.

As shown in Figure 1, several deep conceptual developments lie at the heart of this enterprise [2]. First among these is an appreciation of the importance of symmetries, whether preserved or broken, and dimensionality in shaping the phenomenology of condensed quantum matter [3] [4] [5]. This helped dispel the (lazy) notion that the study of condensed matter is a messy affair replete with details and tedium. Instead, the universality of critical phenomena revealed that disparate quantum systems have common explanations for their phenomenology: forsaking the need for microscopic details, we have learnt to focus on understanding the onset of order that can be quantified using symmetries and dimensionality alone [6]. Remarkable agreement with careful and increasingly sophisticated experiments, conducted over decades and in a plethora of systems, has reinforced this paradigm named after Ginzburg, Landau and Wilson (GLW). Further, it can be argued [7] that quantum mechanical effects are almost completely irrelevant to the physics of a wide variety of emergent states of matter lying within the GLW paradigm; systems of interacting fermions and quantum spins are important outliers and at the heart of our search for quantum matter.

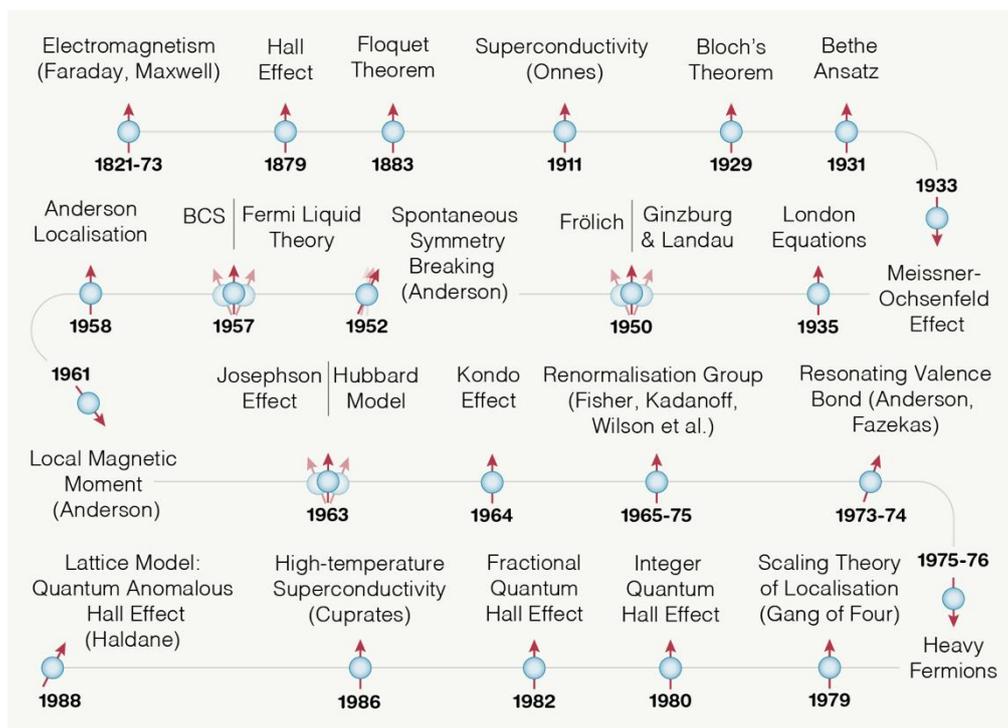

**Figure 1**: A (very rough) map displaying the evolution of our understanding of quantum condensed matter physics across the 19[th] and 20[th] centuries seen through discovered phenomena, ideas, concepts and theories in this field as well as related fields. The choice of events is representative so as to keep the size of the map manageable, and the map should rightfully contain many other important events. The map is strongly influenced by parallel developments in quantum mechanics, statistical mechanics, optics, materials science (and especially electronics), quantum field theory, high energy physics, gravitation & cosmology etc.

These insights were further elevated through the development of several key concepts, such as adiabatic continuity, renormalisation group and effective theories. The first of these helped with understanding how, for instance, the excitation spectrum of an interacting system of electrons can be essentially equivalent (in terms of the quantum numbers) to that of a non-interacting gas of electrons. In this way, Landau revealed the existence of a class of gapless phases of electronic quantum matter known as Fermi liquids that best represent our understanding of conventional metals [8]. BCS superconductivity [9] and the quantum Hall effects [10] are other prominent examples of collective phenomena that are successfully described by effective theories. Further, concomitant developments in statistical mechanics, quantum field theory and many-body theory led to the formulation of an overarching language by which to delineate the theories that describe various phases of matter most effectively, as well as track the passage between them. Known as the renormalisation group, this language revolutionised our understanding of how the interplay of degrees of freedom at various scales of energy, distance and momentum enables a layered structure for emergent phenomena spanning these scales [11].

Experimental advances have pushed the frontier continuously. Today, we seek novel materials that can superconduct at room temperatures and ambient pressures, others with exotic topological excitations that can be used as quantum processors, next-generation semiconductors with unprecedented electron mobilities, multiferroic quantum materials that allow control of their magnetic and charge properties in unconventional ways etc. This has coincided with a shift in perspective on quantum matter beyond the GLW paradigm of spontaneously broken symmetries, emergent local order parameters and collective bosonic excitations. Moving beyond the physics of effectively non-interacting electrons, the importance of several new ingredients in shaping phases of quantum matter has been recognised. As shown in Figure 2 (left panel), this includes strong electronic correlations, frustration (i.e., the interplay of competing inter-particle interactions, lattice geometry or disorder), quantum fluctuation-driven criticality at zero temperature, the topology and (quantum) geometry of wavefunctions and many-particle entanglement. A topical example involves the search for exotic strongly interacting quantum liquid phases of electrons by quenching their kinetic energy in so-called "flat bands" engineered within Moiré lattices of layered van der Waals materials [12] [13].

**Research challenges for the theorists.**

A theoretical understanding of many of these experimental advances remains challenging. The physics of several novel phases of quantum matter are not easily understood in terms of non-interacting quasiparticles emergent from symmetry breaking and whose quantum numbers can be identified. Prominent examples here include gapless metallic states that fall outside Landau's Fermi liquid paradigm [14], and gapped liquid-like insulating states that appear to possess signatures of topological order and lie beyond the GLW paradigm [15]. Many quantum materials show signatures at finite temperatures of quantum criticality [16] linked with the breakdown of ordered insulating states of matter (whether symmetry broken or topologically ordered), and associated with the presence of non-Fermi liquids [14]. In turn, these poorly understood metals sometimes appear to be the parent states for emergent unconventional superconducting states [17]. Even less is known about the roles played by disorder and out-of-equilibrium dynamics in shaping quantum matter, and how the entanglement encoded within their eigenstates relates to their properties.

An overarching theoretical framework for these puzzles is missing. We offer a few more examples. First, the physics associated with the melting of the Mott insulating many-particle state and the formation of a Fermi surface with proximate gapless excitations can seldom be captured from a perturbation-theoretic treatment [18]. Similarly, little insight is available into what causes the Landau quasiparticles of the Fermi liquid (that are adiabatically continuous to the excitations of a non-interacting gas of electrons) to break down, and be replaced by some other form of gapless fermionic excitations near the Fermi surface of a non-Fermi liquid in two spatial dimensions (and above) [19].

Further, only certain features of topological order are known at present [15]. This includes a degeneracy of the ground state that is sensitive to the topology of the spatial manifold on which the system is placed, topological excitations that carry fractional charge and anyonic statistics, the presence of current-carrying edge states and the existence of long-range entanglement (in the form of subsystem entanglement entropy that is dependent on the topological degeneracy). Another outstanding puzzle involves understanding the competition between the tendencies of a system of quantum spins to order magnetically and be screened by interactions with a reservoir of conduction electrons [20]. Much attention has also been drawn recently towards understanding the interplay between strong electronic interactions, band topology and quantum geometry of the Hilbert space in revealing novel states of quantum matter within dispersionless bands [21] [22].

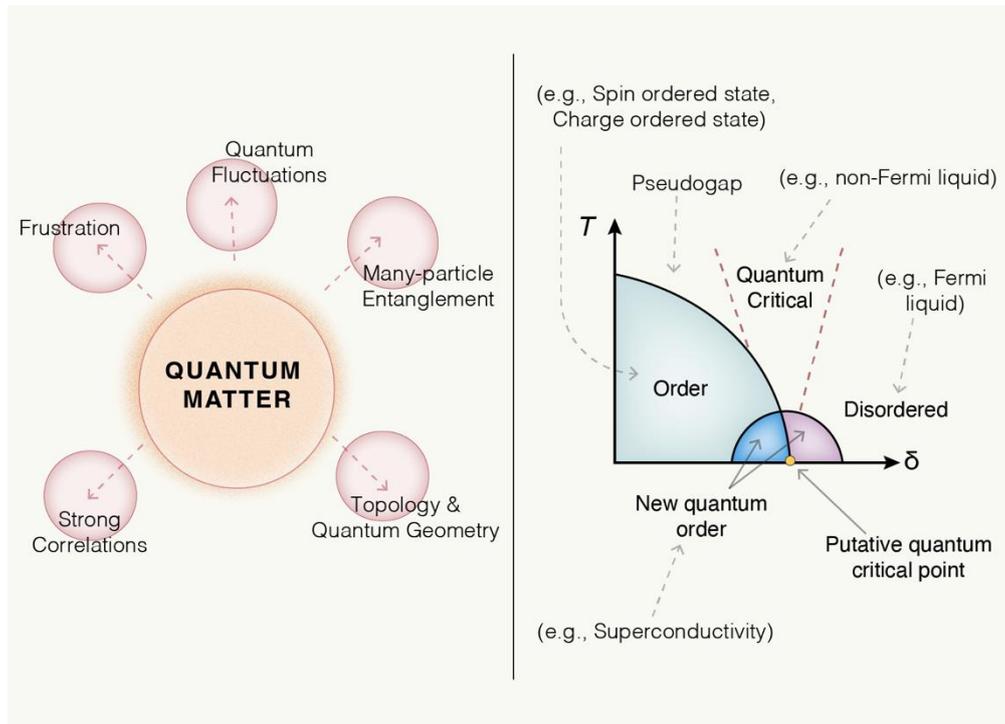

**Figure 2**: (Left Panel) The phenomenology of quantum matter is shaped by several organisational principles, some of which we identify here. (Right Panel) A generic temperature (*T*, y-axis) vs. quantum fluctuation control parameter (δ, x-axis) phase diagram for emergent phenomena in interacting quantum matter involving a quantum phase transition. See text for discussion.

Physicists often capture the essence of their pursuits pictorially in terms of phase diagrams. These figures encapsulate the existence, nature of, and relationships between various phases of matter depending on the external conditions they are subjected to. In the same vein, several of the puzzles related to various forms of quantum matter mentioned above have, through state-of-the-art experimental investigations, been encapsulated into a generic phase diagram shown in Figure 2 (right panel). At its heart lies the physics of a *T=0* quantum phase transition (which may itself correspond to either a point, line or even a phase in the diagram) driven by large quantum fluctuations that destroy an ordered phase, and tuned by some parameter (shown on the x-axis as δ) other than temperature. The ordered state exists at finite temperatures (*T*) and over a certain range of δ, and often correspond to a gapped state involving the ordering of spin or charge degrees of freedom. On the other hand, the disordered state typically corresponds to a gapless state such as a Fermi liquid. Passage between the ordered and disordered phases at finite, non-zero temperatures involves traversing a quantum critical regime where the system becomes effectively gapless but without any evidence for Landau quasiparticle excitations (essentially some kind of non-Fermi liquid metal). A poorly understood pseudogapped phase is sometimes proximate to the quantum critical regime at finite temperatures.

Finally, the quantum critical point can itself sometimes be obscured by an emergent phase of quantum matter; very often, this is observed to be a superconductor (and sometimes with an unconventional order parameter). Some obvious questions that arise include: why is this phase diagram (and its close variants) observed in many quantum material systems? In some other materials (e.g., the cuprate family of high-temperature superconductors) a non-Fermi liquid phase has been well established, but the suspected underlying quantum criticality has not. What leads to a pseudogapped state of matter, characterised by various kinds of fluctuations but no firm notion of order? How is an unconventional form of emergent superconducting order tied in with all of this? Much theoretical effort has yielded only partial answers thus far.

**A call to arms.**

Several powerful numerical methods employed presently in meeting these challenges suffer from a variety of strong limitations that limit the range of their applicability. For instance, exact diagonalisation methods are strongly limited due to the exponential growth in the Hilbert space with increase in system size. Quantum Monte Carlo approaches are often restricted to working at high temperatures due to sign problems while computing the partition function at lower temperature regimes. The density matrix renormalisation group method has proved extremely successful in dealing with systems in one spatial dimension, but is unable to access large systems in two spatial dimensions and beyond. Auxiliary model methods such as dynamical mean-field theory have proved very successful in probing the physics of strongly correlated systems, but often lack physical insight due to the opacity of the self-consistency procedure involved. Other methods have access to only perturbative regimes in various parameters. Indeed, the need of the hour is non-perturbative analytic and numerical methods that do not suffer from these limitations [23]. At a broader level, these methods must also be able to identify the effective theories that describe specific phases of quantum matter observed experimentally at various energy scales. They must also capture the essence of the quantum phase transitions that describe the passage between phases, and will need a careful analysis of the quantum fluctuations that drive these transitions. Further, the construction of novel mathematical formalisms and algorithms for numerical simulations will benefit immensely from guidance obtained from the understanding of first principles calculations of electronic bandstructure, as well as insights offered by quantum simulators (e.g., cold atomic systems [24]) and computations on hybrid classical-quantum platforms (see, e.g., [25]). Quantum simulators, for instance, afford unprecedented precision and control of various aspects of well-known model Hamiltonians. By uncovering the universal principles that guide emergent phenomena in quantum matter, theorists can bridge the distance between the simplicity and intuition that guides their analyses and the inherent complexity of the materials they wish to study.

To sense the difficulty involved in this enterprise, let us consider the case of understanding the family of cuprate superconducting materials [17]. It is safe to say that the phenomenology of these materials corresponds to that of doped Mott insulators that show d-wave superconductivity with a surprisingly high transition temperature at optimal hole doping. While many other features are known experimentally, a coherent and complete theoretical explanation of the essential physics remains largely elusive. Where should we start from? For more than three decades, most efforts have focussed on the Mott physics believed to be contained within the (almost) isolated planes of copper (Cu) and oxygen (O) atoms. Here, a popular effective model to start from is the single-band Hubbard model of electrons with strongly repulsive on-site interactions and nearest-neighbour hopping on a two-dimensional square lattice. The simplicity of the doped 2D Hubbard model (with only the three parameters of inter-site hopping amplitude, on-site repulsion and chemical potential) is a strong allure for obtaining a universal explanation that can explain the physics of an entire family of materials. Alternatively, one could start from the so-called *t-J* model, i.e., a model comprised of nearest-neighbour spin-exchange interactions between localised spin-1/2 moments and the correlated hopping of doped holes that is enforced by a constraint that excludes double-occupancy of any lattice site. Indeed, the *t-J* model can be reached as an effective model from the Hubbard model in the regime of very strong on-site repulsion. Surprisingly, despite concerted efforts, only broad hints at some of the essential features of the cuprates have been glimpsed from analysing these two models.

One may wonder whether a single-band model of strongly correlated electrons can ever yield an overarching understanding of such a complex phenomenon? Or should we use a three-band model for the Cu-O planes that is likely more complete but also surely less tractable? Can a highly accurate first-principles calculation yield an answer to this question? Even if the doped 2D Hubbard model were to be the correct starting point, can it offer broad explanatory power for the phenomenology of the cuprates by helping unveil (in the language of the renormalisation group) effective theories for the various phases of matter observed experimentally in those materials? Can a detailed analysis of these effective theories, in turn, yield accurate results that agree with experimental results? That this has not been possible till now further stresses the importance on the development of non-perturbative methods [23]. While the lack of space precludes a detailed discussion, it is fascinating to note that a host of similar questions arise in several other families of quantum materials. This includes (i) the heavy-fermion materials that contain lattices of magnetic moments screened by conduction electrons, (ii) Moiré systems comprised of strongly correlated electrons in almost dispersionless bands, as well as (iii) multi-orbital systems such as the pnictides and nickelates that display a rich interplay of spin, charge and orbital degrees of freedom.

There are at least two other areas where advances through the creation of broad theoretical frameworks are called for. First, experimental access to out-of-equilibrium dynamics at ultrafast timescales has opened the door to the exploration of physics well beyond the established equilibrium framework that is presently accessible theoretically [26]. Developments in the field of non-equilibrium statistical mechanics can perhaps offer welcome insights on how to proceed. The second open area involves fleshing out the connection between many-particle entanglement and experimentally observable properties of quantum matter [27]. Progress here will not only improve our classification of various emergent forms of matter, but it will also surely offer insight on how to predict and manipulate the functionality of materials appropriate to the creation of novel quantum devices. Further, even as we focus on developing the mathematics enabling the creation of novel methods, it is equally important to bear in mind that they should offer physical insight that can guide our intuition further. Finally, their success must be judged by the accuracy and efficacy in explaining the phenomena at hand, and not simply by notions of mathematical elegance.

**Looking to the Future**

Around a hundred years on from the conclusion of the first quantum revolution, we stand poised on the cusp of another. Even as considerable interest lies in designing quantum devices for sensing, communication and computation, the exploration of quantum matter offers a rich playground for the creation of technologies based on quantum mechanical rules of operation. Some broad ideas for applications that leverage the many-particle entanglement encoded within correlated quantum matter are presented in Figure 3; work has already begun on many of them all over the world. We end by outlining a few broad challenges for future research that we believe will significantly deepen our understanding of quantum matter in general, and are relevant to a large variety of quantum materials.

- First, can we build a unifying framework for understanding how electronic correlations manifest in the interplay of charge, spin and orbital degrees of freedom, and lead thereby to various ordered states of quantum matter, e.g., Mott insulators, itinerant as well as local moment magnetism, superconductors etc.?
- Second, is there a connection between non-Fermi liquid metals and unconventional superconductors? If so, how does knowing this help with understanding the cause of the attraction that leads to pair formation between (otherwise strongly repulsive) electrons?

- Third, can we stabilise the existence of Mott insulating ground states without the spontaneous breaking of any continuous symmetries (e.g., Néel antiferromagnetism)? If so, do they have anything in common with the quantum liquid states actively sought in quantum Heisenberg antiferromagnets placed on geometrically frustrated lattices?
- Fourth, can we build an overarching framework for understanding the emergence of topologically ordered states of matter within systems of strongly correlated electrons populating (almost) flat bands? Can this offer concrete proposals on how to create, detect and manipulate fractionally charged excitations in these systems?
- Fifth, are there universal patterns of many-particle entanglement at the heart of various forms of quantum matter, and can they be directly probed experimentally?
- Finally, how does disorder interplay with electronic correlations in shaping quantum matter, and can this knowledge be useful in the creation of quantum devices?

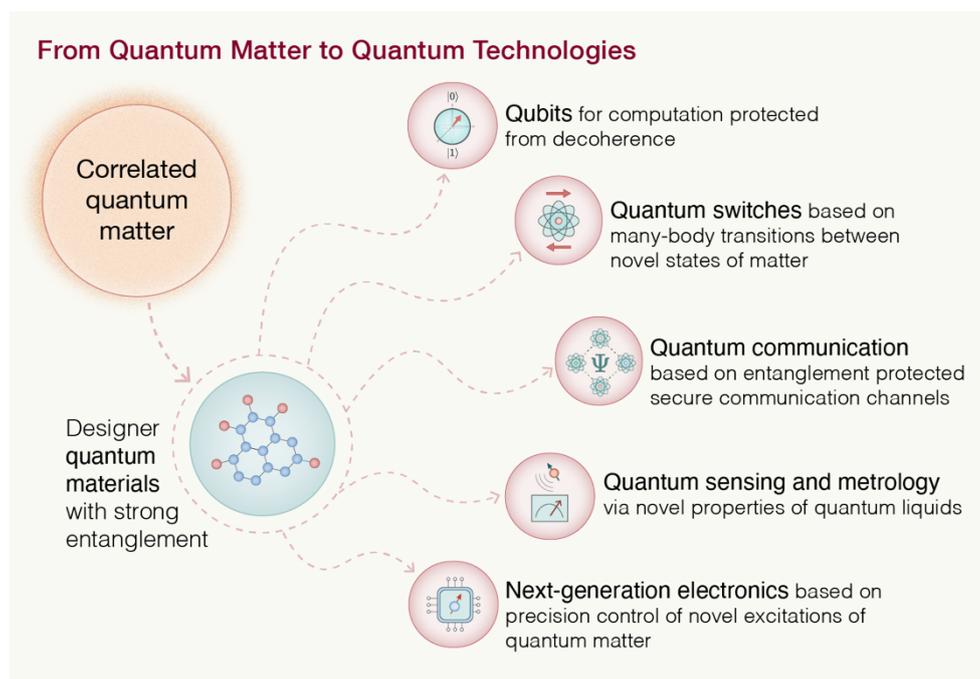

**Figure 3**: Proposals for quantum technologies that leverage the strong entanglement encoded within correlated quantum matter.

The challenges outlined just above as well as elsewhere in this article represent unparalleled opportunities in understanding the organisational principles that govern the physics of quantum matter. Just as liquefying helium enabled the exploration of low-temperature physics and the discovery of superconductivity, efforts in meeting the challenges laid out here will likely enable the prediction of new quantum materials whose properties are tuned to match our expectations of next-generation technologies. It is important to recognise that pathbreaking science must drive the creation of quantum technologies, and that this can only be achieved through the harmonious interplay of experimental achievements and theoretical understanding.

**Acknowledgements**
S.L. thanks the SERB, Govt. of India for support through MATRICS grant MTR/2021/000141 and Core Research Grant CRG/2021/000852. S.L. is deeply grateful to Abhirup Mukherjee, Bhavtosh Bansal, Deepshikha Jaiswal-Nagar, Sumiran Pujari, Arghya Taraphder, Anuradha Bhat, Narayan Banerjee, N. Kamaraju and Sayan Choudhury for insightful comments on the manuscript.

**Author Contribution Statement**

S.L. devised the project and the main conceptual ideas. M.S. contributed to the initial section, covering historic developments displayed in Figure 1. M.S. prepared all the figures in the manuscript. S.L. wrote the manuscript.